\documentclass[twocolumn,twoside]{IEEEtran}

\ifCLASSINFOpdf

\else

\fi

\ifCLASSOPTIONcompsoc
\usepackage[caption=false, font=normalsize, labelfont=sf, textfont=sf]{subfig}
\else
\usepackage[caption=false, font=normalsize]{subfig}
\fi
\usepackage{lipsum}%
\usepackage[dvipsnames]{xcolor}

\usepackage{balance}
\usepackage{multicol}   % for equalize of last page columns
\usepackage{cite}
\usepackage{gensymb}
\usepackage{multirow}
\usepackage{graphics}
\usepackage{epsfig}
\usepackage{graphicx}
\usepackage{epstopdf}
\usepackage{textcomp}
\usepackage{amsmath}
\usepackage{mathtools}
\usepackage{filecontents}
\usepackage{lipsum,color}
\usepackage{amssymb}
\usepackage{float}

\usepackage{amsthm}

\usepackage{braket}

\usepackage{times} % assumes new font selection scheme installed
\usepackage{amsthm}  % for proof 
\usepackage{amsfonts}

\theoremstyle{break}
\begin{document}
\title{Balancing Decentralized Trust and Physical Evidence: A Blockchain–Physical Layer Co-Design for Real-Time 3D Prioritization in Disaster Zones}

\author{Mohammad~Taghi~Dabiri,~Mazen~Hasna,~{\it Senior Member,~IEEE},
	~Aiman~Erbad,~{\it Senior Member,~IEEE}, \\
	~Rula~Ammuri,~and~Khalid~Qaraqe,~{\it Senior Member,~IEEE}
		\thanks{M.T. Dabiri is with the College of Science and Engineering, Hamad Bin Khalifa University, Doha, Qatar. email: (mdabiri@hbku.edu.qa).}
		
	\thanks{Mazen Hasna and  Aiman Erbad are with the Department of Electrical Engineering, Qatar University, Doha, Qatar (e-mail: hasna@qu.edu.qa; aerbad@qu.edu.qa).}
	
		\thanks{R. Ammuri is with Professionals for Smart Technology (PST), Amman, Jordan (email: rammuri@pst.jo).}
		
	\thanks{Khalid A. Qaraqe is with the College of Science and Engineering, Hamad Bin Khalifa University, Doha, Qatar, and also with the Department of Electrical Engineering, Texas A\&M University at Qatar, Doha, Qatar (e-mail: kqaraqe@hbku.edu.qa).}
	\thanks{This publication was made possible by NPRP14C-0909-210008 from the Qatar Research, Development and Innovation (QRDI) Fund (a member of Qatar Foundation), Texas A\&M University at Qatar, and Hamad Bin Khalifa University, which supported this publication. }
	}

% make the title area
\maketitle
%\vspace{-1cm}
%%%%%%%%%%%%%%%%%%%%%%%%%%%%%%%%%%%%%%%%%%%%%%%%%%%%%%%%%%
%%%%%%%%%%%%%%%%%%%%%%%%%%%%%%%%%%%%%%%%%%%%%%%%%%%%%%%%%%
\begin{abstract}
During disaster response, making rapid and well-informed decisions about which areas require immediate attention can save lives. However, current coordination models often struggle with unreliable data, intentional misinformation, and the breakdown of critical communication infrastructure.
A decentralized, vote-based blockchain model offers a compelling substrate for achieving this real-time, trusted coordination.
This article explores a blockchain-driven approach to rapidly update a dynamic 3D crisis map based on inputs from users and local sensors. Each node submits a timestamped and geotagged vote to a public ledger, enabling agencies to visualize needs as they emerge. However, ensuring the physical authenticity of these claims demands more than cryptography alone.
We propose a dual-layer architecture where mobile UAV verifiers perform physical-layer attestation and issue independent location flags to the blockchain. This dual-signature mechanism fuses immutable digital records with sensory-grounded trust. 
We analyze core technical and human centric challenges, ranging from spoofing and vote ambiguity to verifier compromise and connectivity loss, and outline layered mitigation strategies and future research directions.
As a concrete instantiation, we present a UAV mapping scheme leveraging modulated retro-reflector (MRR) sensors and 3D-aware LoS placement to maximize verifiability under urban occlusion, offering a path toward resilient, trust-anchored crisis coordination.
\end{abstract}


\IEEEpeerreviewmaketitle

%% ---------------------------------------
%% ---------------------------------------

%\tableofcontents
%%%%%%%%%%%%%%%%%%%%%%%%%%%%%%%%%%%%%%%%%%%%%%%%%%%%%%%%%%%%
\section{Introduction}
In disaster scenarios rapid decisions on where and how to deploy limited resources are life-critical. Rescue teams face a complex dilemma: almost all affected areas need urgent help, but response capacity is limited. Effective prioritization becomes essential, not just to act, but to act smart. Whether restoring power, dispatching medical aid, or reopening communication lines, agencies must determine \textit{which zones} demand \textit{which services} first. This demands a live, accurate, and trustworthy map of ground-level needs. %, ideally co-generated by people, devices, and sensors already present in the area.

%While satellite imaging and UAV reconnaissance offer top-down visibility, they often lack contextual granularity. In contrast, bottom-up inputs, submitted directly by those affected, can reveal localized urgencies: injured individuals, infrastructure damage, or lack of access. A compelling model is \textbf{crowd-prioritization}, where citizens and edge devices vote on local needs and severity levels. If these inputs are geotagged, timestamped, and immutable, they can help agencies construct a \textit{crisis heatmap}, enabling adaptive and just-in-time allocation of resources.

While satellite imaging and UAV reconnaissance offer top-down visibility, they often lack the fine-grained contextual detail needed for rapid decision-making in disaster zones~\cite{americanredcross2025challenges, kucharczyk2021remote}. 
In contrast, bottom up inputs submitted directly by those affected can capture localized urgencies, such as injured individuals, infrastructure damage, or blocked access routes, that are often invisible in coarse resolution imagery. A compelling model is \textbf{crowd-prioritization}, where citizens and edge devices vote on local needs and severity levels. When geotagged, timestamped, and immutably stored, these inputs enable agencies to construct a \textit{crisis heatmap} for adaptive, just-in-time resource allocation.

Blockchain-based voting mechanisms appear ideal: decentralized, tamper-proof, auditable, and capable of handling asynchronous inputs without relying on central control \cite{vladucu2023voting}. %In such a model, each vote acts as a verified claim of \"need\" that is (i) geotagged to a specific location, (ii) timestamped to capture temporal urgency, and (iii) semantically labeled to indicate the nature and severity of the request. 
In such a model, each vote acts as a verified claim of \emph{need}, consisting of:
\begin{itemize}
	\item \textbf{Location:} Geotagged to a specific position on the map.
	\item \textbf{Time:} Timestamped to reflect temporal urgency.
	\item \textbf{Semantic Label:} Describing the type and severity of the request (e.g., medical, power, access).
\end{itemize}
Agencies can then apply domain specific weighting to these votes, for example giving higher priority to medical requests over power requests, and combine them with top down observations from satellites or UAVs. This multimodal integration enables the generation of high-resolution, context-rich crisis maps, allowing for faster and more informed decision-making on resource allocation.

However, this vision carries significant risks. Blockchain alone cannot confirm whether a vote truly comes from the physical location it claims, or whether the submitting entity is a legitimate user, a sensor, or a bot. Connectivity can be intermittent or entirely lost in disaster zones. GPS signals may be spoofed, jammed, or degraded due to environmental factors \cite{rustamov2023improving}. Furthermore, votes may be submitted by actors located outside the disaster zone, exploiting the openness of the network and overwhelming the decision logic. These threats erode the integrity of the system and expose the fragility of purely digital trust models in hostile conditions.

Hence, location trust becomes the core bottleneck. A blockchain-based voting layer, while suitable for immutable and decentralized data sharing, lacks an intrinsic mechanism to bind information to physical presence. Without anchoring votes to the actual geography of the disaster zone, the system risks strategic manipulation or misrepresentation. This limitation calls for a rethinking of the architectural separation between communication and physical verification. While blockchain emphasizes trust \textit{without} centralized oversight, physical layer mechanisms such as line of sight verification, signal propagation, or hardware based interaction focus on trust \textit{with} direct sensory proof. In our scenario, these seemingly opposite philosophies can become complementary: physical-layer verification enforces grounded presence, while blockchain ensures global auditability and tamper-resistance. 
Together, they can provide a resilient pathway toward trusted, crowd sourced prioritization during crises, an approach that this article aims to conceptualize, analyze, and illustrate through system level design.

To this aim, this article proposes a layered architecture that combines public blockchain infrastructure with physical-layer verification by mobile UAVs. User-submitted votes are digitally signed, geotagged, and semantically labeled, while UAVs perform independent location verification and attach trust flags via separate blockchain transactions. This dual-signature model enhances transparency, filters manipulation, and allows relief agencies to prioritize needs based on verifiable context. We describe the system components, identify key technical and human-centric challenges, propose mitigation strategies, and present a case study illustrating UAV-based coverage optimization and retro-reflector-assisted optical interrogation in a 3D urban setting. Together, these contributions pave the way toward resilient, trust-grounded crisis coordination systems.

\section{Literature Review}
The convergence of blockchain and wireless communication is increasingly seen as key to embedding trust, privacy, and resilience into next-generation networks. Surveys such as Hafeez \textit{et al.}~\cite{10182294} and Ghourab \textit{et al.}~\cite{ghourab2023interplay} highlight blockchain’s role in decentralized UAV security and the value of physical-layer features (e.g., CSI, RSS) to reinforce cryptographic trust.
Beyond surveys, recent works combine blockchain with physical-layer mechanisms for resource coordination and secure exchanges: Luo \textit{et al.}~\cite{luo2025convergence} employ backscatter-assisted links for ultra-low-power 6G IoT transactions, and Cheng \textit{et al.}~\cite{cheng2023blockchain} extend blockchain-based settlement to space–air–ground networks. Other efforts focus on data anchoring for UAV sensing without verifying reported positions~\cite{adu2023digital}.
In disaster response, blockchain has mainly supported accountability and UAV coordination rather than real-time, crowd-driven situational awareness. Examples include BIM integration for post-disaster permitting~\cite{nawari2019blockchain}, blockchain-based UAV swarm coordination with hybrid DPoS–PBFT consensus~\cite{hafeez2024blockchain}, and privacy-preserving UAV location authentication via zero-knowledge proofs~\cite{pan2024privacy}. These works, while advancing system level trust, do not physically verify ground user reports or integrate independent UAV attestations into a shared, tamper proof crisis map, gaps that are addressed by our proposed blockchain physical layer co design.

\textbf{Synthesis and research gap:} Prior art collectively establishes (i) blockchain as a credible substrate for decentralized coordination and tamper-evident logging across UAV/IoT systems~\cite{10182294,luo2025convergence,cheng2023blockchain}, (ii) the role of physical-layer features in strengthening trust beyond pure cryptography~\cite{ghourab2023interplay}, and (iii) early blockchain-enabled disaster-networking for UAV swarms~\cite{hafeez2024blockchain} as well as privacy-preserving \emph{UAV} location authentication~\cite{pan2024privacy}. 
However, these efforts either anchor system to system interactions (for example UAV to UAV communication, resource trading, or BIM workflows) or validate the UAVs’ own positions, without completing the essential step of verifying civilian ground level reports that drive prioritization during a crisis. In particular, existing frameworks do not fuse \emph{crowd-submitted needs} with \emph{independent, physical-layer attestations} by mobile verifiers, nor do they operationalize multi-verifier (redundancy-aware) confirmations into a public ledger to produce \emph{trust-weighted, real-time 3D crisis maps}.

\textbf{Our contribution} directly addresses this gap by co-designing a blockchain layer (for immutable, auditable vote logging) with a physical verification layer (UAV-based interrogation and attestation). Through dual-signature records (user + verifier), redundancy-aware LoS placement in 3D urban geometry, and low-power optical backscatter via MRRs, the proposed architecture \emph{anchors digital claims to physical reality} and enables actionable, trust-weighted prioritization in disaster zones, a capability not provided by prior literature~\cite{10182294,ghourab2023interplay,luo2025convergence,cheng2023blockchain,adu2023digital,nawari2019blockchain,hafeez2024blockchain,pan2024privacy}.

%%%%%%%%%%%%%%%%%%%%%%%%%%%%%%%%%%%%%%%%%%%%%%%%%%%%%%%%%%%%%%%%
%%%%%%%%%%%%%%%%%%%%%%%%%%%%%%%%%%%%%%%%%%%%%%%%%%%%%%%%%%%%%%%% VERSUS P_T
\begin{figure}
	\begin{center}
		\includegraphics[width=3.35 in]{PP_fig/main_diagram.pdf}
		\caption{System architecture for trusted, crowd-sourced crisis prioritization. Users broadcast signed need reports to a public blockchain, which stores immutable records. UAVs perform physical-layer verification and append location flags (verified / unverified / unknown). Relief agencies extract, weight, and fuse this data to generate actionable response maps.}
		\label{nv1}
	\end{center}
\end{figure}
%%%%%%%%%%%%%%%%%%%%%%%%%%%%%%%%%%%%%%%%%%%%%%%%%%%%%%%%%%%%%%%%
%%%%%%%%%%%%%%%%%%%%%%%%%%%%%%%%%%%%%%%%%%%%%%%%%%%%%%%%%%%%%%%%

%% -------------------------------------------------------------
\section{System Vision}
In our proposed system, each user or sensor node within the disaster zone participates by submitting a signed message to a public blockchain. 
This message contains a location claim, derived from onboard GPS, manual input, or local sensing, along with a semantic label describing the observed need, such as urgent medical assistance, suspicious activity reports, trapped individuals, communication blackout, lack of power, or gas leakage. These submissions are immutable and globally visible, enabling decentralized prioritization without a central server.

To guard against location spoofing or external manipulation, the system integrates an adaptive physical-layer verification step. 
Mobile verifiers, typically UAVs, probe the location claim using \textbf{multi-modal interrogation: high-precision optical links for infrastructure nodes, or passive RF sensing (e.g., Angle-of-Arrival) for standard smartphones.} Depending on signal integrity, line-of-sight constraints, and time-of-flight analysis, each verifier submits one of the following to the blockchain:
\begin{itemize}
	\item \textbf{Verified:} Location and identity of the sender confirmed.
	\item \textbf{Unverified:} The claimed location was scanned, but no corresponding response or user presence was detected.
	\item \textbf{Unknown:} No direct measurement possible (e.g., obstructed LoS).
\end{itemize}

These verifications are submitted as separate blockchain transactions, each digitally signed by the verifying entity (e.g., UAV). This dual layer approach, consisting of the message from the user and the assessment from the verifier, produces a distributed ledger that contains not only the votes but also contextual confidence for each one.

\textbf{Trust scoring emerges as a key mechanism:} agencies can prioritize votes based on the \textbf{verification tier—assigning higher confidence to optically-anchored claims compared to RF-verified ones—}depending on their application domain (e.g., health, power, rescue logistics). Importantly, this model supports heterogeneous strategies across decision-makers, while sharing a common, tamper-proof data substrate.

To support trust-based crisis decision-making across heterogeneous agencies, the system builds on a public blockchain augmented with physical-layer verification. The structure is illustrated in Fig.~\ref{nv1} and consists of the following core components:
\begin{itemize}
	\item \textbf{Open Infrastructure:} A public, permissionless blockchain (e.g., Ethereum-compatible) ensures universal visibility and interoperability among heterogeneous relief agencies.
	
	\item \textbf{Dual Signatures:} Each vote carries two cryptographic signatures:
	\begin{itemize}
		\item \textbf{User:} Claims authorship and local need.
		\item \textbf{Verifier (e.g., UAV):} Confirms or contests user presence at the claimed location.
	\end{itemize}
	
	\item \textbf{Flexible Attestations:} Votes may receive 0, 1, or multiple verifier endorsements. Agencies can weigh votes based on the number and quality of confirmations.
	
	\item \textbf{Why It Matters:} Blockchain records \textit{what} was said, while physical-layer verification adds \textit{where} it was said from, bridging digital inputs with physical reality.
\end{itemize}

%% ------------------------------------------
%% ------------------------------------------
%% ------------------------------------------
\section{Challenges to Grounded Trust}
Despite its promise, integrating blockchain-based coordination with physical-layer verification in disaster zones raises multiple technical, environmental, and cognitive challenges. We group these obstacles across three layers: trust integrity, infrastructure fragility, and human-system interaction.

\subsection{Trust and Adversarial Integrity}
While decentralization improves resilience, it also dilutes accountability. In our model, both users and verifiers contribute to the trust fabric, but either can potentially become an adversarial agent. This section outlines core threats to vote integrity from malicious behavior or compromised components.

\subsubsection{Sybil Attacks and Identity Misuse}
Open participation invites Sybil attacks, where an adversary floods the network with fake or duplicated identities. Without secure identity binding or uniqueness constraints, vote maps can be skewed, overwhelming genuine requests.

\subsubsection{Location Spoofing and Geographic Misinformation}
Users may falsely claim physical presence in high-priority areas. Even when digital signatures are valid, geographic assertions remain unverifiable without physical-layer checks, making spoofing a core risk in location-based systems. While optical links are naturally resistant to remote spoofing, RF-based verification faces specific threats from AI-driven signal synthesis and SDR-based replay attacks, necessitating robust channel profiling.

\subsubsection{Compromised Verifiers or Censoring Behavior}
Verifiers such as UAVs serve as trust anchors, but they too can be compromised. A malicious or externally tampered verifier may suppress, forge, or selectively issue location flags. Such censorship attacks distort crisis perception by excluding legitimate needs from downstream analysis. Since verifiers carry disproportionate weight, their integrity must be auditable, cross-validated, or even made subject to quorum-based verification.

\subsubsection{Replay or Relay Attacks}
Attackers may capture legitimate votes and rebroadcast them elsewhere to mislead the system. Without context-aware timing and physical-layer attestation, such replayed or relayed claims can poison trust metrics and lead to misallocated resources.

%----------------------------------------
\subsection{Infrastructure and Temporal Fragility}

\subsubsection{Connectivity Blackouts and Communication Asymmetry}
In disaster zones, traditional communication infrastructure may be partially or completely unavailable. Users may lack access to cellular networks, Internet, or even reliable local links. This leads to asymmetric communication: while UAVs can transmit from above, ground users may struggle to broadcast votes in real-time.

\subsubsection{Time-Varying Ground Truth vs. Immutable Logs}
Blockchain’s immutability ensures accountability, but may conflict with the dynamic nature of real-world crises. A user’s need may change quickly, for example escaping from danger, receiving aid, or losing connectivity, while the recorded vote remains static. This temporal mismatch can lead to misinformed prioritization if downstream systems rely on outdated or context-inappropriate data.

\subsection{Human-System Interaction and Cognitive Constraints}

\subsubsection{Human-in-the-Loop Uncertainty}
While decentralized voting empowers citizens to shape crisis response, it also introduces a less predictable actor: the human. In high-stress disaster scenarios, users may lack the training, composure, or contextual clarity to generate semantically accurate reports. Misclassification of needs, vague labeling, or even accidental location claims can introduce semantic noise into the system.

\subsubsection{Ambiguity and Usability Gaps}
The success of any blockchain-based participation model hinges on the usability of its interface. If mobile apps are too complex, language-specific, or reliant on consistent network feedback, users may fail to complete submissions or abandon them altogether. Involving civilians with diverse digital literacy and linguistic backgrounds introduces further ambiguity. Unlike signal errors, human-generated uncertainty is harder to detect or filter, yet equally capable of skewing prioritization.

\subsection{Privacy and Decision Fusion}

\subsubsection{Privacy vs. Interoperability}
Revealing exact location and need type may expose vulnerable users to surveillance or stigmatization. Yet agencies must access interpretable data across domains. Balancing anonymity with coordination is essential.

\subsubsection{Weighting and Conflict Resolution}
Votes may be contradictory, unverifiable, or ambiguous. Relief teams must devise domain-specific heuristics or machine learning tools to resolve conflicts while preserving fairness and response efficiency.

\begin{figure*}
	\begin{center}
		\includegraphics[width=7 in]{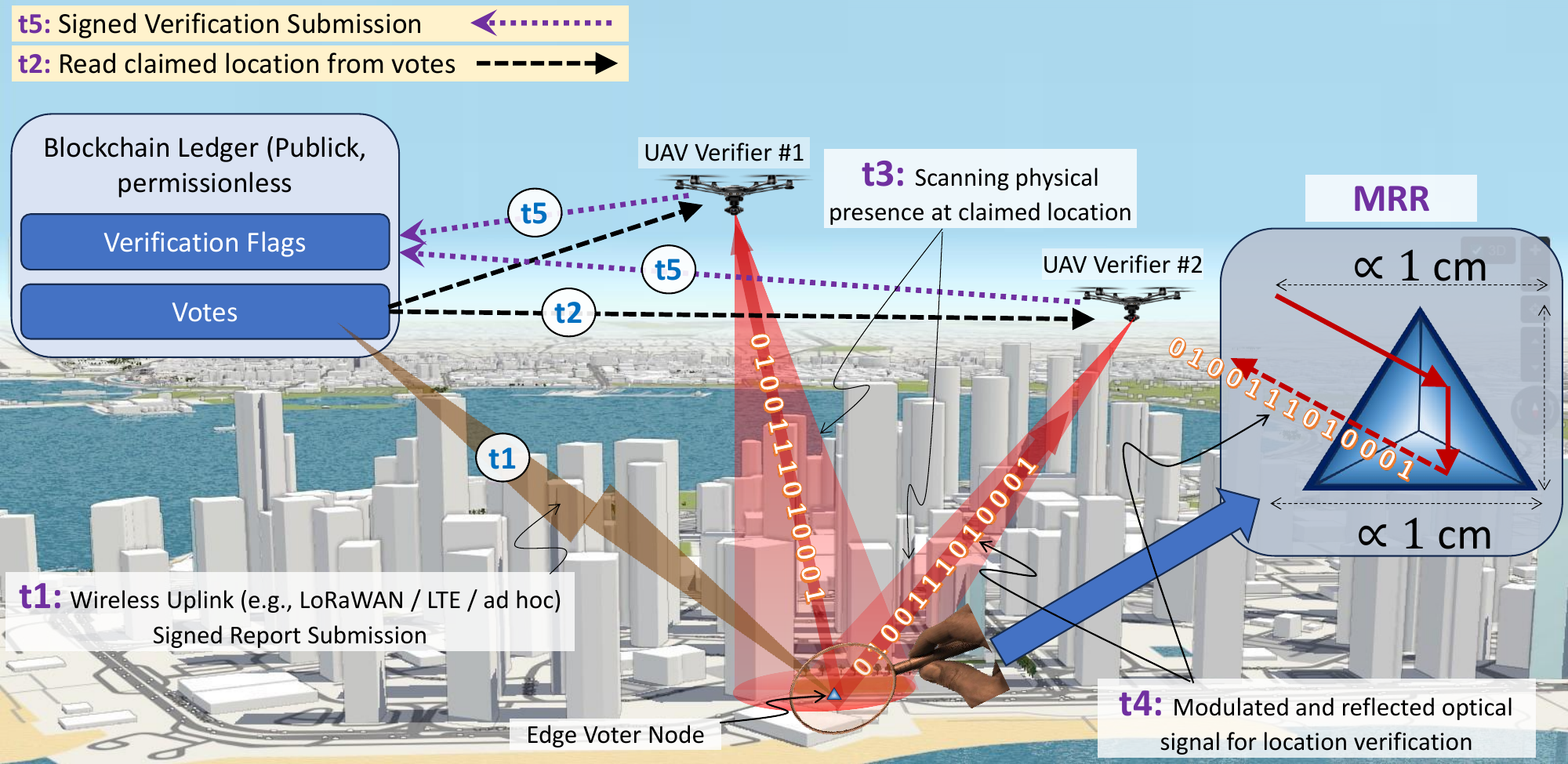}
		\caption{UAV placement for LoS coverage amid complex 3D urban obstructions, ensuring that each ground location is visible to multiple verifiers for resilient physical-layer attestation. The system operates in five stages: (t1) vote submission from ground users; (t2) blockchain-based retrieval of location claims by UAVs; (t3) directional interrogation using optical or mmWave beams; (t4) response detection via MRRs; and (t5) signed verification logging on the public blockchain. MRRs enable energy-efficient, alignment-tolerant feedback suitable for constrained environments. Notably, in certain scenarios, UAVs in phase (t3) may directly verify reported claims by scanning the area for environmental cues such as thermal signatures, motion patterns, or acoustic anomalies, enabling non-cooperative validation when user devices are offline or silent.
		}
		
		\label{nv2}
	\end{center}
\end{figure*}

%% ---------------------------------------------------------
%% ---------------------------------------------------------
\section{Solutions and Future Directions}
Building on the challenges outlined in Section~III, we present corresponding solution pathways grounded in physical-layer innovation, decentralized coordination, and human-centric design. 
Each strategy targets a core vulnerability, whether adversarial, infrastructural, or cognitive, while also highlighting open research directions for resilient crisis coordination.

\subsection{Trust and Adversarial Integrity}
Establishing robust trust in open, crisis-prone environments requires more than cryptographic validity. When both users and verifiers are potential adversaries, safeguarding vote integrity demands deeper architectural thinking. Below, we outline key research directions to strengthen trust anchors in decentralized, physical-layer-aware systems:

\begin{itemize}
	\item \textbf{Redundancy as Anti-Censorship:} Relying on a single verifier exposes the system to targeted manipulation. Defining and optimizing LoS redundancy thresholds ($N_{\text{LoS}} > 1$) offers a natural defense against UAV compromise, flag suppression, and single-point failure.
	
	\item \textbf{Verifier Reputation Dynamics:} 
	Designing transparent on chain reputation systems for verifiers, based on accuracy, consistency, and community feedback, can help detect bias, censorship, or coordinated misinformation over time.
	
	\item \textbf{Secure Hardware Anchors:} Incorporating tamper-proof modules (e.g., TPM, TEE, or PUF-based attestation) into UAV verifiers remains an open challenge, especially under SWaP (size, weight, and power) constraints typical of disaster UAVs.
	
	\item \textbf{Quorum-Based Verification Models:} Exploring consensus-like mechanisms among UAVs (e.g., threshold signing or cross-verification before flag finalization) may enhance robustness against flag forgery or verifier collusion.
	
	\item \textbf{Adversarial Simulation Frameworks:} There is a growing need for simulation environments that model verifier compromise, Sybil flooding, and dynamic vote manipulation to evaluate system-level resilience before deployment.
\end{itemize}

\subsection{Physical Constraints and Environmental Fragility}
Environmental uncertainty and degraded infrastructure challenge reliable physical-layer verification. The following directions highlight pathways to resilient sensing under operational constraints:
\begin{itemize}
	\item \textbf{Risk-Aware Scanning:} UAVs can prioritize flight trajectories based on spatial entropy or anomaly density, reducing coverage latency in high-risk zones.
	
	\item \textbf{Multi-Modal Verification:} Combining optical, mmWave, and RF backscatter modalities improves LoS robustness and extends coverage to obstructed or cluttered areas.
	
	\item \textbf{Passive Presence Estimation:} 
	In blackout scenarios, UAVs may infer user presence through thermal signatures, passive RF emissions, or residual motion patterns, enabling non cooperative verification.
\end{itemize}

\subsection{Human Factors and Usability Constraints}
User-facing components must remain effective under stress, uncertainty, and diverse digital literacy. Key directions include:
\begin{itemize}
	\item \textbf{Guided and Accessible Reporting:} Interfaces should employ iconography, multilingual support, and context-aware geotagging to minimize cognitive load during crisis reporting.
	
	\item \textbf{Participation Feedback Loops:} UAVs can emit lightweight acknowledgment signals (e.g., visual or RF beacons) to confirm vote receipt and verification, reducing user anxiety and submission redundancy.
\end{itemize}

\subsection{Semantic Clarity and AI-Augmented Prioritization}
To enhance operational clarity and decision quality, semantic tools and AI-assisted analytics can help bridge the gap between raw crisis data and actionable insight:
\begin{itemize}
	\item \textbf{User-Centric Semantic Layers:} Designing vote interfaces and flag reports using crisis-relevant ontologies (e.g., shelter, injury, supply need) ensures consistency and interpretability across diverse literacy levels and languages. Embedding iconographic and language-independent cues can reduce reporting ambiguity under stress.
	
	\item \textbf{AI-Driven Flag Aggregation:} 
	Leveraging machine learning to cluster and classify user votes based on content, location, and historical context can help agencies prioritize and manage responses more effectively. Such models can detect anomalous patterns, emerging risks, or underreported regions in real-time.
	
	\item \textbf{Multi-Source Fusion with Satellite Context:} Integrating UAV-ground data with satellite imagery (e.g., flood extent, thermal mapping, road blockages) enables more accurate identification of high-priority zones. This multimodal fusion supports more strategic deployment of limited relief assets.
\end{itemize}

\subsection{System-Level Design Opportunities}
Scalable deployment of trusted verification systems requires coordination, observability, and empirical benchmarking. Promising directions include:
\begin{itemize}
	\item \textbf{Swarm-Based Coordination:} UAVs can adopt decentralized consensus or local gossip protocols to optimize scan coverage, reduce redundancy, and maintain fault tolerance without central control.
	
	\item \textbf{Transparent Ground Truth Interfaces:} Public dashboards exposing vote density, flag confidence, and conflict zones enable participatory oversight and multi-agency coordination.
	
	\item \textbf{Emulation-Driven Evaluation:} Reference testbeds combining blockchain middleware with UAV and user emulators are essential for assessing system-level tradeoffs in latency, coverage, and adversarial resilience.
\end{itemize}

\begin{figure}
	\begin{center}
		\includegraphics[width=3.35 in]{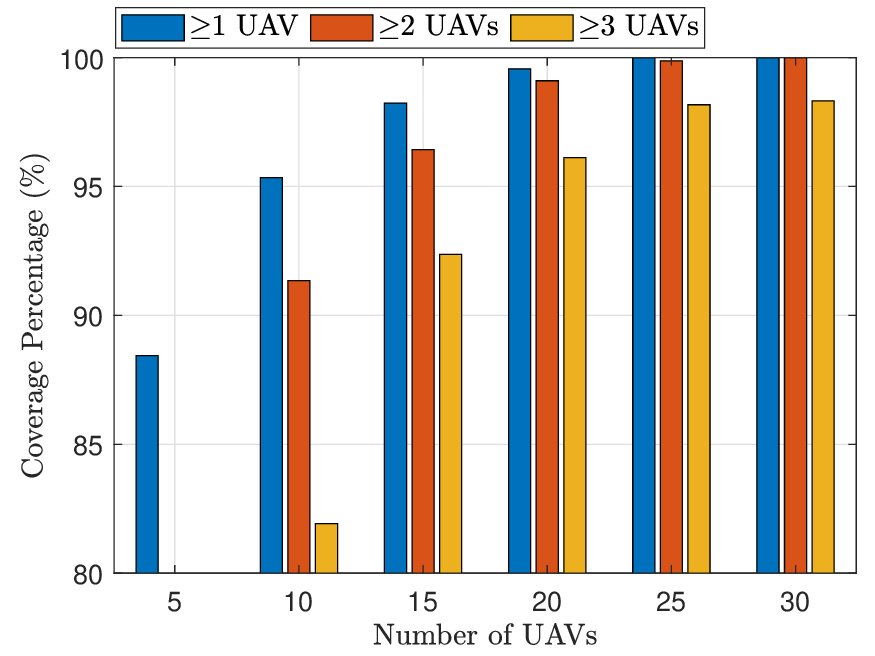}
		\caption{Percentage of covered area vs. number of UAV verifiers under different minimum LoS redundancy constraints ($N_{\text{LoS}} = 1, 2, 3$). Higher redundancy improves resilience but requires more UAVs.}
		\label{nv3}
	\end{center}
\end{figure}

\begin{figure}
	\begin{center}
		\includegraphics[width=3.35 in]{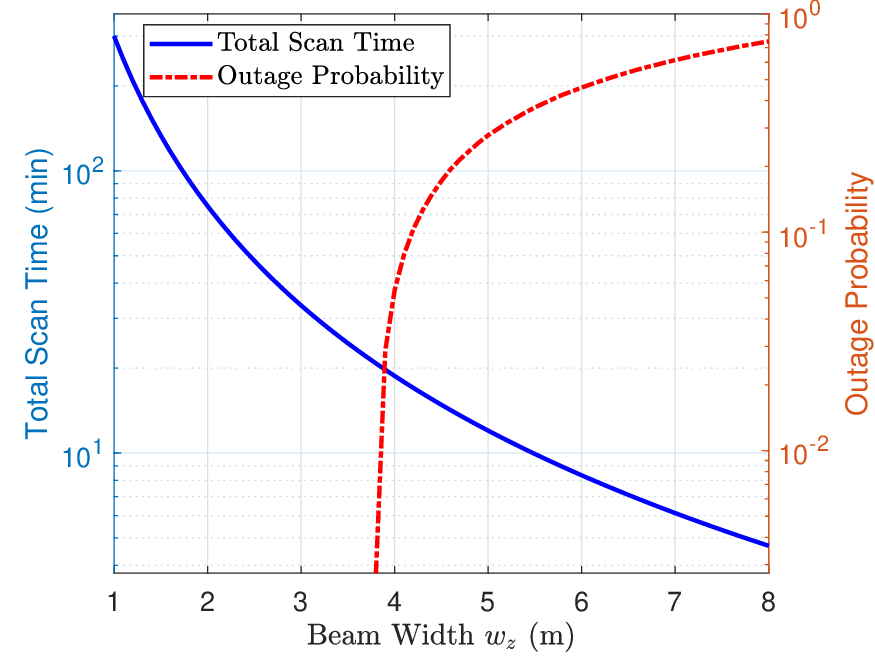}
		\caption{Tradeoff between scanning time and outage probability as a function of optical beamwidth $w_z$. Wider beams reduce scan duration but lower received power at the MRR, increasing communication failure risk.}
		\label{nv4}
	\end{center}
\end{figure}

%% -------------------------------- Modified
%% -------------------------------------------
\section{A Case Study: Optical Scanning for Physical-Layer Verification}
In this section, we illustrate a concrete realization of our architecture using UAV–assisted optical verification in dense urban environments. 
% START OF CHANGE: Added scoping sentence
\textbf{While our proposed architecture supports multi-modal verification (Tier-1 Optical and Tier-2 RF), this case study focuses specifically on the optical Tier-1 implementation.} This is because optical LoS alignment in 3D obstructed environments represents the more challenging constraint compared to omnidirectional RF coverage, which has been extensively studied in prior literature. 
% END OF CHANGE
The goal is to enable accurate physical-layer attestation even in GPS-degraded or occluded areas, leveraging compact, low-power sensors mounted on ground users and optimized UAV positioning above a real 3D cityscape.

\subsection{Time-Slotted Operation Overview}
To illustrate the end-to-end interaction between user submissions, verifier actions, and physical-layer attestation, we divide the system's operational timeline into five conceptual time slots, as illustrated in Fig.~\ref{nv2}:
\begin{itemize}
	\item \textbf{t1 – Vote Submission:}  
	An entity located within the disaster zone, such as a user device or an autonomous sensor, generates a vote containing geolocation, timestamp, and semantic labels describing the observed need. This message is digitally signed and transmitted via a wireless uplink to a public blockchain.
	
	\item \textbf{t2 – Query Acquisition by Verifiers:}  
	Verifier agents (e.g., UAVs) periodically scan the blockchain to retrieve new location claims submitted within their coverage region. Based on these entries, they compute candidate target coordinates for physical verification.
	
	% START OF CHANGE: Updated t3 to reflect Optical + RF
	\item \textbf{t3 – Directed Interrogation (Optical/RF):}  
	Each verifier initiates a scan toward the reported location. For Tier-1 nodes, this involves a tightly focused optical beam; for Tier-2 users, the UAV utilizes directional RF sensing (e.g., AoA). This scanning phase aims to interrogate the spatial claim and establish line-of-sight (LoS) interaction.	
	% END OF CHANGE
	
	\item \textbf{t4 – Response Capture and Evaluation:}  
	If the target location contains a cooperative responder (e.g., a passive optical retro-reflector or active transponder), the verifier detects and decodes the returned signal. This roundtrip interaction provides strong physical-layer evidence of presence and synchronization.
	
	\item \textbf{t5 – Blockchain Logging of Verification Outcome:}  
	The verifier then publishes a digitally signed verification flag—classified as \emph{verified}, \emph{unverified}, or \emph{unknown}—back to the blockchain. This metadata becomes permanently attached to the original vote, enabling downstream trust weighting and conflict resolution.
\end{itemize}
This time-slotted architecture enables asynchronous coordination across distributed agents while preserving strong spatial and temporal anchoring for each vote.

While the {\bf t3} step is often associated with direct interaction, it can also operate independently of explicit user response. In certain scenarios—such as heat signatures from fires, motion clusters near trapped zones, or localized acoustic cues—the verifier may assess the plausibility of reported needs through environmental sensing alone. This enables non-cooperative validation, particularly when user devices are damaged, offline, or deliberately silent. Importantly, such passive interrogation opens new design dimensions: defining scenario-specific sensing models, adaptive confidence metrics, and task-specific hardware capabilities. These directions highlight a rich, largely untapped space for resilient verification under constrained or adversarial conditions.

\subsection{3D-Aware UAV Positioning for LoS Coverage}
Timely and reliable physical-layer verification relies critically on the ability of UAVs to establish LoS connectivity with user-reported coordinates. In dense urban environments, where signal occlusion and multipath effects are prevalent, ensuring LoS is essential for enabling focused optical interrogation or directional RF probing. The availability of LoS not only increases the success rate of verification but also reduces UAV scanning latency, energy consumption, and response time during critical events.

To systematically enable this, UAV placement must be informed by 3D structural data of the environment. By leveraging digital urban models, UAVs can be positioned to maximize visibility across the disaster zone while ensuring robustness against individual verifier failure \cite{dabiri2025joint}. Specifically, we define a minimum LoS redundancy level, $N_{\text{LoS}}$, which denotes the required number of independent UAVs that must have unobstructed visibility to each ground point. Setting $N_{\text{LoS}} > 1$ is crucial not only for mitigating signal blockage and hardware failure, but also for increasing resilience against adversarial compromise, such as UAV hacking, selective censorship, or false negative reports. Redundant verification paths ensure that no single point of failure can suppress or distort physical-layer validation.

As a representative example, we consider a \(3 \times 3\) km\(^2\) region in the West Bay district of Doha, Qatar, an area characterized by high-rise buildings and complex urban geometry. The 3D building specifications used in this case study were extracted from publicly available OpenStreetMap data~\cite{osm}. The objective is to identify the minimal UAV deployment set that guarantees maximal LoS coverage under different redundancy constraints. Using a greedy optimization algorithm, we iteratively place UAVs to satisfy LoS coverage for all user points under $N_{\text{LoS}} = 1, 2, 3$ requirements.

Fig.~\ref{nv3} illustrates the tradeoff between the number of deployed UAVs and the resulting LoS coverage area. As expected, higher redundancy demands a greater number of UAVs to maintain full visibility, especially in occlusion-heavy zones. These results underscore the importance of 3D-aware, redundancy-constrained deployment strategies for scalable and robust physical verification in urban disaster settings.

\subsection{High-Precision Optical Verification via VeCSEL Scanning and MRR Links}
Emerging laser sources such as vertical-cavity surface-emitting lasers (VeCSELs) offer compact, energy-efficient, and directionally controllable optical beams, making them highly suitable for fine-grained 3D scanning in disaster zones. With high modulation rates and beam-shaping capabilities, VeCSEL-equipped UAVs can efficiently interrogate spatial claims under LoS conditions, enabling both physical validation and secure communication initiation.

Complementing this, we consider the use of modulated retro-reflectors (MRRs) as lightweight, passive responders on the ground. An MRR consists of three mutually orthogonal mirrors and a micro-scale optical modulator that encodes binary data by switching a reflected laser signal. These devices require no internal power for reflection and consume only microwatts for modulation, making them ideal for emergency deployment \cite{10944637}. 

The directional nature of the optical link enables centimeter-level localization and is inherently immune to GPS spoofing, jamming, or signal relay attacks. However, due to the MRR’s limited aperture, communication reliability depends on beam alignment. This introduces a tradeoff: wider beams ($w_z$) cover larger areas faster but reduce received power, increasing outage probability.

To study this effect, we simulate a GPS-denied urban scenario (Fig.~\ref{nv3}) where 20 UAVs collaboratively scan the 3D area using VeCSEL beams. Each UAV sweeps its assigned region with beamwidth $w_z$, dedicating 50 ms per pointing direction. Fig.~\ref{nv4} presents the impact of $w_z$ on total scan duration and link reliability. As expected, larger $w_z$ accelerates coverage but increases link failure rates. Optimal operation requires balancing speed and communication quality based on mission context.

In summary, combining VeCSEL-based scanning with MRR links enables robust, secure, and low-power location verification under GPS-denied conditions.

%% -----------------------------------------------
\section{Conclusion}

Disaster time decision making depends on the rapid aggregation of trustworthy, location aware information from unstructured environments. While blockchain enables decentralized and auditable record-keeping, it cannot independently guarantee physical presence. This article presented a hybrid architecture that fuses blockchain-based voting with physical-layer verification using UAVs. By assigning verifiable location flags to user-submitted votes, the system enhances trust without relying on centralized control.

We explored the challenges inherent in such fusion—ranging from Sybil attacks and connectivity blackouts to cognitive uncertainty and verifier compromise—and outlined technical pathways to address them. A specialized case study using MRRs and 3D-aware UAV positioning in dense urban settings illustrated the feasibility and tradeoffs of real-world deployment.

Ultimately, grounded trust requires more than secure data—it demands anchoring digital claims to physical reality in dynamic, adversarial, and resource-limited environments. We envision future systems where physical sensing, secure computation, and human-in-the-loop design co-evolve to enable resilient, community-driven crisis response infrastructures.

%%%%%%%%%%%%%%%%%%%%%%%%%%%%%%%%%%%%%%%%%%%%%%%%%%%%%%%%%%%%%%
%%%%%%%%%%%%%%%%%%%%%%%%%%%%%%%%%%%%%%%%%%%%%%%%%%%%%%%%%%%%%%
\bibliographystyle{IEEEtran}
\balance
\bibliography{IEEEabrv,myref}

\end{document}